\documentclass{article}

\PassOptionsToPackage{numbers, compress}{natbib}

\usepackage[preprint]{neurips_2025}

\usepackage[utf8]{inputenc} 
\usepackage[T1]{fontenc}    
\usepackage{hyperref}       
\usepackage{url}            
\usepackage{booktabs}       
\usepackage{amsfonts}       
\usepackage{nicefrac}       
\usepackage{microtype}      
\usepackage[dvipsnames]{xcolor}
\usepackage{capt-of}
\usepackage{amsmath,amssymb,amsthm}

\usepackage{dblfloatfix}
\usepackage{graphicx}
\usepackage{subcaption}
\usepackage{enumitem}
\usepackage{tikz}
\usepackage{braket}
\usepackage{yquant}
\usepackage{tikzscale}

\usetikzlibrary{chains}
\usetikzlibrary{quotes}

\definecolor{drawioBlue}{HTML}{C6DBFA}
\newcommand{\quirk}{QuIRK }
\newcommand{\quirks}{QuIRKs }
\newcommand{\dr}{DR }

\newcommand{\R}{\mathbb{R}}

\newcommand{\M}{\mathcal{M}}

\makeatletter
\DeclareRobustCommand{\rvdots}{%
  \vbox{
    \baselineskip5\p@\lineskiplimit\z@
    \kern-\p@
    \hbox{.}\hbox{.}\hbox{.}
  }}
\makeatother
\yquantdefinebox{vdots}[inner sep=0pt]{$\vdots$}

\title{\quirk{}: Quantum-Inspired Re-uploading KAN}

\author{%
  Vinayak Sharma \\
  Arizona State University, Tempe \\
  \texttt{Vinayak.Sharma@asu.edu} \\
  \And
  Ashish Padhy  \\
  National Institute of Technology, Rourkela \\
  \texttt{ashishpadhy1729@gmail.com} \\
  \And
  Lord Sen  \\
  National Institute of Technology, Rourkela \\
  \texttt{123cs0226@nitrkl.ac.in} \\
  \And
  Vijay Jagdish Karanjkar  \\
  Arizona State University, Tempe \\
  \texttt{vkaranjk@asu.edu} \\
  \And
  Sourav Behera  \\
  National Institute of Technology, Rourkela \\
  \texttt{souravdipu2002@gmail.com} \\
  \And
  Shyamapada Mukherjee  \\
  National Institute of Technology, Rourkela \\
  \texttt{mukherjees@nitrkl.ac.in} \\
  \And 
  Aviral Shrivastava \\
  Arizona State University, Tempe \\
  \texttt{Aviral.Shrivastava@asu.edu} \\
}

\begin{document}
\maketitle

\begin{abstract}
  Kolmogorov-Arnold Networks or KANs have shown the ability to outperform classical Deep Neural Networks, while using far fewer trainable parameters for regression problems on scientific domains. Even more powerful has been their interpretability due to their structure being composed of univariate B-Spline functions. This enables us to derive closed-form equations from trained KANs for a wide range of problems. This paper introduces a quantum-inspired variant of the KAN based on Quantum Data Re-uploading~(\dr) models. The Quantum-Inspired Re-uploading KAN or \quirk model replaces B-Splines with single-qubit \dr models as the univariate function approximator, allowing them to match or outperform traditional KANs while using even fewer parameters. This is especially apparent in the case of periodic functions. Additionally, since the model utilizes only single-qubit circuits, it remains classically tractable to simulate with straightforward GPU acceleration. Finally, we also demonstrate that \quirk retains the interpretability advantages and the ability to produce closed-form solutions. 
\end{abstract}

\section{Introduction} \label{section:Introduction}
Kolmogorov-Arnold Networks (KANs) \cite{KAN_2024} were introduced as a class of networks based on the Kolmogorov-Arnold Representation Theorem (KART) \cite{arnold2009representation,kolmogorov1957representation} and promised to be more efficient in smaller function learning tasks than Deep Neural Networks (DNNs). According to KART, a multivariate continuous function $f$ on a bounded domain can be written as a finite composition of univariate functions.

\begin{equation} \label{eq:KART}
    f(x_1,x_2, \hdots , x_n) = \sum_{q=1}^{2n+1} \Phi_q \left( \sum_{p=1}^{n} \phi_{q,p}(x_p) \right)
\end{equation}

Where $f: [0,1]^n \rightarrow \R$ defines the outer function and $ \phi_{q,p} : [0,1] \rightarrow \R$ and $\Phi_q : \R \rightarrow \R$ define the inner functions used for approximation. A major challenge when applying this theorem to machine learning (ML) is that the 1-D functions $\phi_{q,p}$ might be non-smooth or even fractal. However, \citeauthor{KAN_2024} argued that in the average case most functions in science are smooth and have sparse composite structure. This allows us to use smooth functions, in their case B-Splines, as the basis to create a network that can scale to arbitrary breadth and depths.

The choice of B-Splines as the basis for KANs has proven to be effective, but, they are not the only functions we may use in order to construct KAN models. In principle, we could use any univariate universal function approximator as the basis for a KAN model. However, for practical purposes we would like these univariate functions to be expressive, compact and easy to compute. Data Re-Uploading (DR) circuits \cite{Data_Reuploader_2020} are a class of QML algorithms that fit these characteristics and hence can form a promising new basis for KAN models.

Quantum Machine Learning (QML), is a field of machine learning which seeks to exploit the special properties of quantum circuits in order to create new more efficient ML models. \citeauthor{Data_Reuploader_2020}, introduced the Data Re-Uploading Classifier, which was proven to be a Universal Classifier based on the \emph{Universal Approximation Theorem} \cite{hornik1989multilayer} of DNNs (Eqn. \ref{eq:UAT}).

\begin{equation} \label{eq:UAT}
    h(\vec{x}) = \sum_{i=1}^{N} \alpha_i \cdot \sigma(\vec{w_i} \cdot \vec{x} + b_i) \; \quad ; \alpha_i, b_i \in \R
\end{equation}

Specifically the authors used a corollary which stated that $\sigma$ could be a non-constant finite linear combination of periodic functions. This allowed DR models to create universal function approximators using rotational quantum gates. Critically, this proof also makes DR models eligible univariate functions to utilize as building blocks for KANs.

We utilize the universal approximation functionality of DR models in order to construct a new variant of the KAN model called \quirks. \quirk incorporate DR models in place of B-Splines while maintaining the same scalability in depths and breadth structure as KANs. Due to the inherent higher dimensionality of qubits  compared to real-valued univariate functions, DR models are far more expressively compact. This increased dimensionality is due to two main factors - (1) The vectorized nature of qubits caused by superposition. (2) The complex-valued nature of qubit states. This allows our DR based \quirk to use fewer trainable parameters and smaller model sizes when compared to KANs for a similar level of accuracy as shown experimentally in the results (\ref{section:Results}) section.

Another important aspect of \quirks is that they can be easily accelerated on GPUs. We refer to our model as `\emph{Quantum Inspired}' because, although we utilize a QML model, the structure of our network does not require any quantum computers for training or inference. Due to the lack of entanglement, all our circuits are factorizable into single qubit circuits and can be computed using $2\times 2$ matrix multiplications (matmuls) on simulators. These matmul operations are far more optimized for modern GPUs as compared to B-Splines used by KANs. This allows for better computational scaling as compared to KANs. Additionally, the use of these quantum inspired mathematical structures allows us to leverage the expanded feature space of  quantum systems. It is important to note that current backpropagation implementations and optimizers struggle with complex numbers and hence we leverage the capabilities of quantum simulators in this domain to train our models.

Therefore, the main features of the \quirk model can be summarized as follows:
\begin{enumerate}
    \item  Data re-uploading based scalable network architecture,
    \item Parameter efficient KAN model leveraging the higher dimensionality of quantum systems,
    \item Computationally scalable due to simple $2 \times 2$ matrix multiplication based implementation.
\end{enumerate}

\section{Related Work} \label{section:Related_Work}
The idea of a Quantum Kolmogorov-Arnold Network (QKAN) was first proposed by \citet{ivashkov2024qkanquantumkolmogorovarnoldnetworks}. In their work they used linear combination of Chevyshev polynomials, through QSVT \cite{QSVT}, as learnable activation functions in place of B-splines. However, the focus of their work was creating a theoretical framework to implement KAN on quantum computer and their number of qubits scale linearly with inputs while gates scale quadratically with layers leading to an exponential increase in computational cost. We propose a novel architecture that combines the benefits of the Kolmogorov-Arnold Network (KAN) and the Data Re-Uploading (DR) Classifier to create a scalable, quantum-inspired, classical architecture, while still using a predetermined number of qubits making it NISQ compatible.

Kolmogorov-Arnold Network (KAN) \cite{KAN_2024} are networks based on the Kolmogorov-Arnold Representation Theorem (KART) \cite{arnold2009representation}. As opposed to Multi-Layer Perceptrons (MLPs) which are composed of units with fixed non-linear functions of weighted sums of inputs, KAN units are sums of trainable univariate functions (or activations). In the original work these activations were B-Splines \cite{curry1947spline}. The original KART was defined for 2 layers, \citet{KAN_2024} generalized the KART to multiple layers which allowed them to construct scalable network architectures.

Our work explores replacing B-Splines used by Liu and Wang et al. with a universal classifier similar to the work by \citet{shen2023neuralnetworkarchitecturewidth}. While they explored using DNNs as trainable activation functions, we propose using Data Re-Uploading (DR) classifiers \cite{Data_Reuploader_2020} as the activation functions. These models have the same approximation guarantees via the UAT as DNNs while being significantly cheaper to compute by exploiting the larger feature space of quantum systems.

Data Re-uploading classifiers (DR) \cite{Data_Reuploader_2020} are a class of QML models that exploit hybrid classical-quantum computation to circumvent the no-cloning theorem. \citet{Data_Reuploader_2020} proposed an analogue to classical DNNs which learn multiple functions in a single variable to approximate a target function. The DR circuit therefore `\emph{re-uploads}' the input Data at the beginning of each layer, allowing the model to learn multiple functions in a single variable. This approach allowed the authors to establish approximation guarantees for DR classifiers using the Universal Approximation Theorem (UAT) \cite{hornik1989multilayer}.

A key advantage of this approach is the fact that DR models are universal function approximators in a single variable, which allows us to use them as extremely efficient activation functions in the KAN architecture. A single qubit can be simulated very efficiently on a classical computer via $2 \times 2$ matrix multiplication, thus making our architecture extremely scalable.

\section{\quirk Architecture} \label{section:architecture}
\begin{figure}[htbp]
    \centering
    \includegraphics[width=0.89\textwidth]{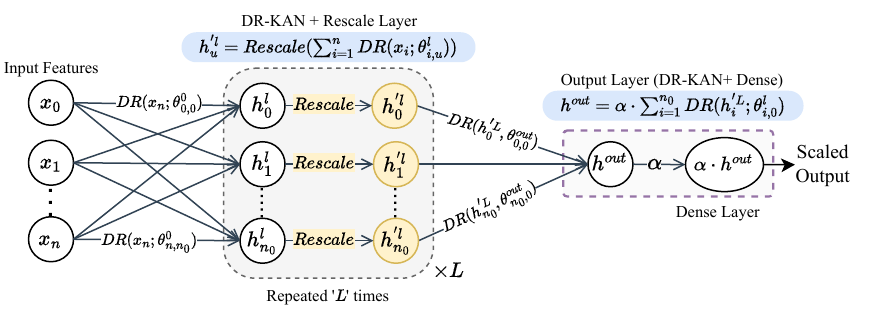}
    \caption{\quirk Macro architecture. Each incoming edge to a \quirk layer is a single qubit DR circuit. The output of each hidden \quirk layer ($h^l_u$) is then rescaled to the range $[0,\pi]$ creating node $h^{'l}_u$. The output of the network is a \quirk layer with a single unit followed by an \emph{optional} dense layer which scales the output to the desired range.}
    \label{fig:qkan_macro}
\end{figure}

\quirk largely inherits the structure of classical KAN models. As such, it is composed by an Inner and Outer function as shown in Eqn. \ref{eq:KART}. The outer function ($\Phi$) refers to the operation of summation of the \dr circuit outputs in each layer. The inner functions ($\phi$) are \dr circuits which learn univariate functions. Additionally, we introduce a `\emph{Rescale}' layer between \quirk layers in order to account for the domain and range of \dr circuits. As shown in Fig. \ref{fig:qkan_macro}, we can create deeper models by stacking blocks of \quirk layer and rescale layers. Finally, a classical dense layer with a single unit is optionally appended to scale the output data ($[-1,1] \rightarrow \R$). We can therefore distill the \quirk model into 4 main components:

\begin{enumerate}
    \item \textbf{\quirk layers,} which consist of multiple \quirk units.
    \item \textbf{\quirk units,} which each consist of sums of \dr circuits over the input data.
    \item \textbf{Rescale layers,} which rescale the output of each \quirk unit from $\R \rightarrow [0,\pi]$.
    \item \textbf{Classical Dense layers,} which can optionally be appended to the output of the \quirk layer to scale the output $[-1,1] \rightarrow \R$ if required.
\end{enumerate}

\begin{figure}[hbp]
    \centering
    \begin{tikzpicture}
        \begin{yquant}
            qubit {$q : \ket{0}$} q;
            [this subcircuit box style={draw, dashed,"DR Layer", inner ysep=7pt}]
            subcircuit {
                    qubit {} q;
                    box {$\phi(x)$} (q);
                    box {$U_0(\Theta^{0})$} (q);
                } (q);
            text {$\cdots \text{L layers} \cdots$} q;
            box {$\phi(x)$} (q);
            box {$U_L(\Theta^{L})$} (q);
            measure q;
            output {$\M_{Z} = \braket{U(x;\theta)\ket{0}}$} q;
        \end{yquant}
    \end{tikzpicture}
    \caption{Single qubit Data Re-Uploading circuit with `$L$' layers. Each layer consists of a data encoding function $\phi(x)$ following by a trainable unitary $U(\Theta^{i})$. The final measurement is a function of the input `$x$' and the trainable parameters `$\theta = \{\Theta^i | \;\; \forall i \in [0,L]\}$'. Each activation is composed of multiple DR layers.}
    \label{fig:dr_circuit}
\end{figure}
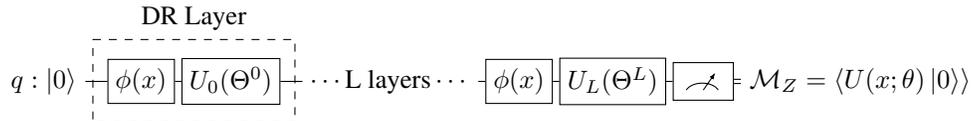

\subsection{\quirk layers} \label{subsubsec:qkan_layers}
\quirk Layers are composed of units which sum over the expectation values of DR circuits. Formally, we can define the output of the $i^{th}$ unit of the layer `$l$' as:
\begin{equation}\label{eq:qkan_unit_abstract}
    h^{(l)}_{i} = \sum_{i=1}^{|I|} DR(x_i;\theta^{l}_{i,u})  \text{ for } x_i \in I
\end{equation}

Where, $I$ is the set of inputs to the layer,$DR(x_i;\theta^{l}_{i,u})$ the DR circuit better the $i^{th}$ unit of the previous layer and $u^{th}$ unit of the current layer with parameters $\theta^{l}_{i,u}$, and `$x$' is the input to the \quirk layer.

We know that the output of the DR circuit is the expectation value of the final quantum state. This state is created by applying a series of parameterized unitary gates which can be combined to into a single unitary ($U(x_i;\theta^{(l)}_i)$) that is applied to the initial state $\ket{0}$. The expectation value of this final state w.r.t the computational basis is represented by the operation $\braket{\psi}$, where $\psi = U(x_i;\theta^{(l)}_i)\ket{0}$. Therefore, we can unpack the DR layer in Equation. \ref{eq:qkan_unit_abstract} as -

\begin{equation}  \label{eq:qkan_unit}
    h^{(l)}_{i}  = \sum_{i=1}^{|I|} \braket{U(x_i;\theta^{(l)}_i)\ket{0}} \text{ for } x_i \in I
\end{equation}

A lot of the power of these layers can be associated to the representative power of \dr models. While in the single qubit setting they may be limited when trying to approximate multi-variate functions, they excel in the univariate scenario. In principle DR models map the 1-D data onto a higher dimensional Hilbert space, a space with 2 complex dimensions in the single qubit space, where they learn a linear map. This linear map is then projected back to the 1-D space. As shown by \citet{boser1992training}, linear maps in higher dimensional spaces can approximate non-linear functions in lower dimensional spaces. This means that the \quirk units can learn arbitrarily complex univariate functions. This idea is further explored in the next section.

\subsection{DR Activations} \label{subsubsec:dr_activations}
Data Re-Uploading models exploit the ability to `\emph{re-upload}' data in hybrid classical-quantum models in order to create universal function approximators. They accomplish via multiple DR-layers each consisting of a data encoding function $\phi(x)$ followed by a trainable unitary $U(\theta_{i})$ as shown in Fig. \ref{fig:dr_circuit}. Each layer can therefore be consolidated into a single unitary $U^{(i)}(x, \theta^{(i)})$ which is a new function of the input `$x$'. The final state of each DR node can therefore be described as:

\begin{equation} \label{eq:dr_node}
    U(x;\theta)\ket{0} = \left(\prod_{i=0}^{L} U^{(L-i)}(x, \theta^{(L-i)})\right) \ket{0}
\end{equation}

The choice of $U(\theta_{i})$ is not fixed and can be constructed using an arbitrary combination of single qubit parameterized gates (ex. $R_Y,R_X$ and $R_Z$). This choice determines the number of trainable parameters used by DR unit and the expressivity of the DR layers. In general $U(\theta_i)=R_Z(\theta_i[0])R_Y(\theta_i[1])R_Z(\theta_i[2])$, which is proven to be the universal SU2 group. Therefore, using DR models introduces 2 hyperparameters to the \quirk model - (1). The number of DR layers in each node, and (2). The choice of $U(\theta_{i})$.

In our implementation, $\phi(x) = R_Y(x)$ and $U(\theta_i) = R_Z(\theta_i[0]) R_X(\theta_i[1])$. Therefore, we use only 2 parameters per DR layer.

\subsubsection{Multi Qubit Extensions} \label{subsubsec:multi_qubit}

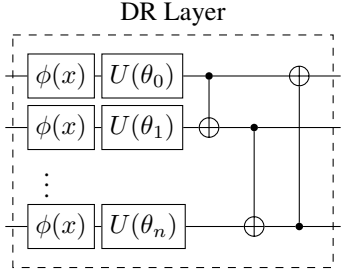
\begin{figure}[h]
    \centering
    \begin{tikzpicture}
        \begin{yquant}
            qubit {} q[2];
            nobit nb;
            qubit {} out;
            [this subcircuit box style={draw,dashed, inner ysep=7pt, "DR Layer"}]
            subcircuit {
                    qubit {} q[2];
                    nobit nb;
                    qubit {} out;
                    box {$\phi(x)$} q;
                    box {$\phi(x)$} out;
                    vdots nb;
                    box {$U(\theta_0)$} q[0];
                    box {$U(\theta_1)$} q[1];
                    box {$U(\theta_n)$} out;
                    cnot q[1] | q[0];
                    cnot out | q[1] ;
                    cnot q[0] | out;
                } (q,nb,out);
        \end{yquant}
    \end{tikzpicture}
    \caption{Multi qubit (n) Data Re-Uploading layer. The layer consists of a set of trainable gates, an input embedding function and an entanglement pattern.}
    \label{fig:dr_multiq_circuit}
\end{figure}

The \quirk model can also be scaled to use multiple qubits per edge in order to learn richer features at the cost of increased computation and more training parameters. The use of multiple qubits per edge also allows us to utilize entanglement. As pointed out by \citet{Data_Reuploader_2020}, entanglement can significantly reduce the number of trainable parameters required to approximate a function. This increased parameter efficiency comes at the risk of an exponential blow up in the size of our matrix multiplications and hence computation. However, due to the overall the structure of \quirk, the DR units only need to learn univariate functions. Therefore, in practice the number of qubits remains below 5 which can be computed via $32 \times 32$ matrices. This is still a reasonable computational cost for most modern computers.

The structure of a multi qubit DR models is outlined in Fig. \ref{fig:dr_multiq_circuit}. The layer consists of a set of trainable gates, an input embedding function and an entanglement pattern. Each qubit in encoded with the same data point, therefore even with multiple qubits, we are still learning a univariate function. The entanglement pattern is used to create a richer feature space as well as have our parallelly learned functions exchange information. The entanglement pattern is not fixed and can be chosen based on the problem at hand.

The multi-qubit \quirk model therefore introduces 2 additional hyperparameters - (1). The number of qubits per edge, and (2). The entanglement pattern between qubits.

\subsubsection{DR circuit Domain and Range}
When using DR models, we have to keep 2 main limitations in mind and design our model around them. The first is that the output of each DR circuit is limited to the range $[-1,1]$. The summation of multiple DR models in a unit therefore expands the range to $[-n,n]$, where `$n$' is the number of DR models in the unit. This plays into the 2nd limitation - the input to our DR models must be in the range $[0,\pi]$. This is due to the periodic nature of the gates used for data encoding, where data shifted by $2\pi$ maps to the same quantum state. These limitations can be addressed by using the `rescale' layer as discussed in Section \ref{subsec:rescale_layer}.

\subsection{Rescale Layer} \label{subsec:rescale_layer}
The use of DR circuits introduces constraints on the input and output ranges of our units. DR circuits encode data using Pauli rotational gates, which are periodic functions and the output for these circuits is a expectation value in the computational basis.

While we can technically have unbounded inputs to a Pauli rotational gate, the periodicity will create an issue where unique data points will map to the same quantum state. As an example consider 2 data points $x,y$ such that $x = y + 4\pi$. Using the $R_Y$ gate to encode the data starting from the $\ket{0}$ state will yield the states $\ket{\phi(x)} = \cos(x/2)\ket{0} + \sin(x/2)\ket{1} $ and $\ket{\phi(y)} = \cos(y/2)\ket{0} + \sin(y/2)\ket{1} $ respectively. Applying the relationship between $x$ and $y$, we can see that $\ket{\phi(y)} = \cos(x/2 + 2\pi)\ket{0} + \sin(x/2 + 2\pi)\ket{1}  = \ket{\phi(x)}$. As we can see, the two data points map to the same quantum state. Additionally, even if the values are within the period, the symmetric nature of sin and cosine functions can create cases where the feature map of two otherwise distinct data points have the same magnitude but differ only by sign. To avoid this issue we need to scale down the values to the range $[0,\pi]$.

Expectation values of quantum states are bounded between $[-1,1]$. Therefore, the output of each \quirk unit is bounded to $[-|I|,|I|]$ where `$|I|$' is the number of incoming edges. This can create issues when the output of the \quirk layer is used as input to the next layer due to the input domain of DR circuits. To address this, we introduce a `Rescale' layer to scale our layer outputs to the domain required for the next layer. The output of the Rescale layer is given by:

\begin{equation}\label{eq:rescale}
    \text{Rescale}(x_i) = \frac{x_i + {(1-b)}}{2\times |I|} \times \pi
\end{equation}

As described in Eqn. \ref{eq:rescale}, the Rescale layer divides each unit output by the upper bound ensuring a value between $[-1,1]$. $b$ is a boolean parameter which is set to $1$ when bias is added the output of the DR-KAN, otherwise it is set to $0$ and by adding $1-b$ to this and dividing by $2$ we shift the range to $[0,1]$. Finally, we multiply by $\pi$ to ensure the input to the next layer is in the range $[0,\pi]$ (i.e. the range of unique values for the Pauli rotational gates).

\subsection{Implementation Optimizations} \label{subsubsec:implementation}
\quirk has been implemented using pennylane \cite{asadi2024hybridquantumprogrammingpennylane} and Tensorflow \cite{abadi2016tensorflowlargescalemachinelearning}. However, we have implemented a number of optimizations to improve the performance and scaling of our model. We employ the following strategies in order to create a practical and scalable model:

\begin{enumerate}
    \item \textbf{Parallel Circuit Execution}: As we are using simulators with single qubits, we can efficiently run multiple circuits are a time.
    \item \textbf{Tensor Network Simulations}: When running multiple batches of data across our parallel circuits, we can efficiently accelerate the execution via tensor network simulations. Nvidia's cuTensorNet allows us to run these circuits on GPUs as well.
    \item \textbf{Dynamic Circuit Creation}: We create our circuits dynamically on model instantiation based on input data sizes.
\end{enumerate}

\begin{figure}[h]
    \centering
--    \includegraphics[width=0.5\columnwidth]{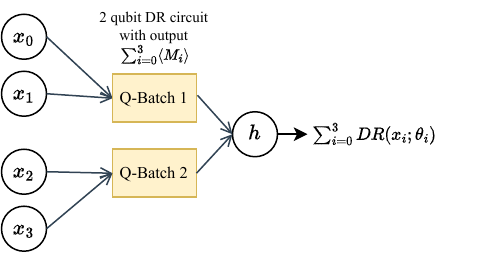}
    \caption{Q-batch execution for 4 dimensional data when `\emph{max\_qubits}'=2. The data is split into 2 batches and 2 circuits are run in parallel.}
    \label{fig:qbatch}
\end{figure}

\subsubsection{Parallel Circuit Execution}
While pennylane already supports batching data when using the `\emph{backprop}' or `\emph{adjoint}' differentiation methods on simulators, that method is limited to batching input data rather than running multiple circuits with different weights as well. In order to achieve parallel execution with different weights we tested 2 separate approaches - (1). Q-batch execution and (2). Threaded Circuit execution.

In Q-batch execution, we define a `\emph{max\_qubits}' parameter which determines the size of our circuit. We then batch our data into batches of size `max\_qubits' and run a single circuit with `max\_qubits' number of qubits. The circuit is run $C=\frac{N}{mq}$ times or `$C$' circuits are run simultaneously, where $N$ is the size of our input and $mq$ is the max\_qubits parameter.

While this method is effective, we have found that overhead in instantiation of these structures presents a bottleneck and as such our overall model runs fastest when using $~5$ qubits.

\begin{figure}[h]
    \centering
    \includegraphics[width=0.5\columnwidth]{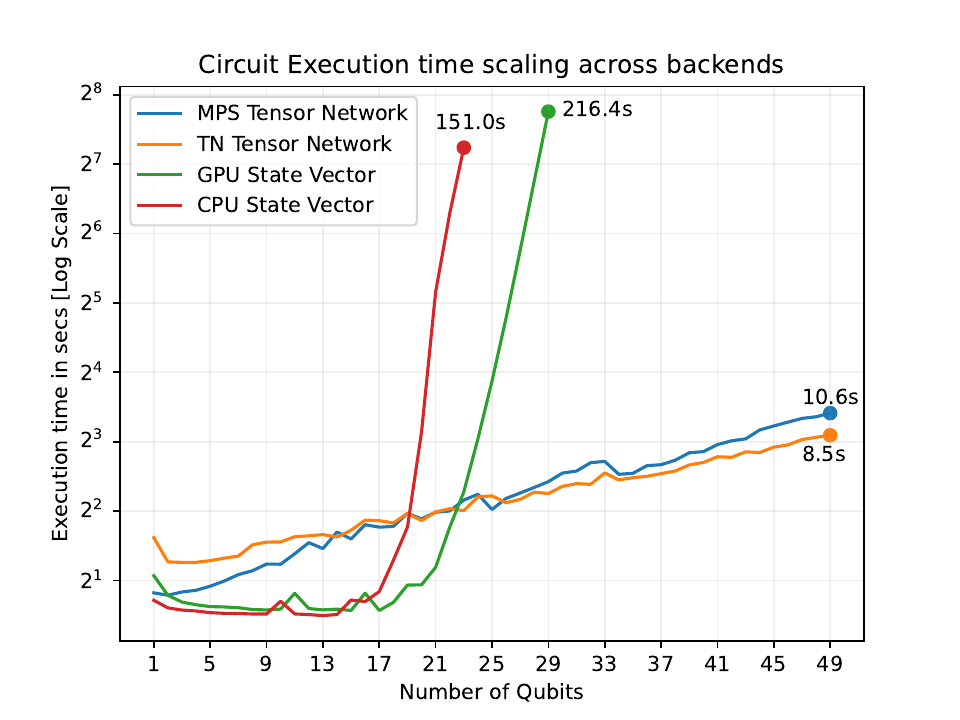}
    \caption{Runtime scaling of with increasing number of qubits across different backends using batch 8- 100 dim data on a 32-thread CPU and A6000 GPU. Tensor Network simulations greatly outperform the state-vector simulator for larger qubit counts but are worse for smaller qubit counts.}
    \label{fig:runtime}
\end{figure}

\subsubsection{Tensor Network execution}
\citet{Markov_2008} showed that we can simulate quantum circuits with $T$ gates and tree-width $d$ (maximum number of qubits entangled together) in $T^{O(1)} e^{O(d)}$ time. As we are running single qubit networks in parallel, we have no entanglement and hence d=1. Therefore, we can simulate our circuits in $T^{O(1)}$ time, which is linear in the number of gates.

Exploitation of tensor network circuits and our unentangled networks allows us to efficiently scale to extremely large ($<30$) qubit networks even on CPU.s. As shown in Fig. \ref{fig:runtime}, we can run a \quirk network with 100 dimensional input, a single \quirk layer with 2 units and a dense layer using a 50-qubit circuit in under 10s on CPU.

\subsubsection{Dynamic Circuit Creation}
Finally, we dynamically create our circuits based on the amount of data required in order to further reduce overhead. This is most relevant when our input data size is not fully divisible by $mq$. In this case, we can dynamically create a smaller circuit with the required number of qubits and add it to the computational graph. This allows us to run our model without any padding or wasted computation and limit the matrix dimensions.

\section{Results} \label{section:results}
\begin{table*}[b]
    \centering
    \caption{\quirk vs Classical KAN on RMSE and Parameter count on Feynman dataset \textbf{[Best in Bold]}}
    \begin{tabular}{@{}l lc lc  lc lc@{}}
        \toprule
        Equation & \multicolumn{2}{c}{\quirk (Ours)} & \multicolumn{2}{c}{\quirk Pruned} & \multicolumn{2}{c}{Classical KAN} & \multicolumn{2}{c}{Classical KAN Pruned}
        \\ \cmidrule(lr){2-3} \cmidrule(lr){4-5} \cmidrule(lr){6-7} \cmidrule(lr){8-9}
                 & Loss                              & \# Params                         & Loss                              & \# Params                                & Loss                          & \# Params    & Loss                 & \# Params \\ \midrule
        I.6.2    & \textbf{8.40$\times 10^{-3}$}     & \textbf{36}                       & $5.95\times 10^{-2}$              & 24                                       & $3.90\times 10^{-1}$          & 136          & $1.51\times 10^{0}$  & 56        \\
        I.15.3x  & \textbf{1.30$\times 10^{-2}$}     & \textbf{36}                       & $1.84\times 10^{-2}$              & 24                                       & $5.14\times 10^{-1}$          & 80           & $6.48\times 10^{-1}$ & 56        \\
        I.26.2   & $7.07\times 10^{-2}$              & 60                                & $2.81\times 10^{-2}$              & 40                                       & \textbf{4.20$\times 10^{-3}$} & \textbf{136} & $1.71\times 10^{-1}$ & 56        \\
        III.9.52 & $3.06\times 10^{-2}$              & 102                               & $4.97\times 10^{-2}$              & 62                                       & \textbf{2.20$\times 10^{-4}$} & \textbf{342} & $8.93\times 10^{-2}$ & 72        \\
        I.44.4   & \textbf{3.74$\times 10^{-2}$}     & \textbf{48}                       & $3.61\times 10^{-2}$              & 24                                       & $4.63\times 10^{-2}$          & 156          & $6.26\times 10^{-1}$ & 56        \\
        \bottomrule
    \end{tabular}
    \label{tab:KAN_vs_qKAN}
\end{table*}

\subsection{Experiments:} \label{section:experiments}
All experiments were run using Pennylane\cite{asadi2024hybridquantumprogrammingpennylane} and Tensorflow \cite{abadi2016tensorflowlargescalemachinelearning}. The system used for the experiments was a single NVIDIA RTX 4080 SUPER GPU and a Ryzen 7 7700X CPU with 32Gb of RAM.

\subsection{Accuracy of \quirk:} \label{subsection:accuracy}
\quirk model are able to showcase strong performance on regression tasks similar to KANs. As seen in table \ref{tab:KAN_vs_qKAN}, \quirk models show RMSE values $~10^{-2}$ on most tasks. We can also see that \quirk outperforms the vanilla KAN model on multiple equations such as $I.6.2$. This can be attributed to the superior performance of \dr models as compared to B-Splines as a unvariate function approximator.

\subsection{Parameter Efficiency of \quirk:} \label{subsection:parameter_efficiency}
\quirk models are able to fit functions using even fewer parameters than classical KANs while maintaining a comparable RMSE loss value. This can be attributed to the efficiency of \dr models over B-Splines in fitting complex functions.

Data Re-Uploading models learn global functions. In contrast, B-Splines learn functions in a piece wise manner and approximate these pieces using basis curves (degree 3 in the case of KANs). This difference can yield significant differences in the parameter efficiency of DR models and B-Splines.

\begin{figure}[t]
    \centering
    \includegraphics[width=\textwidth]{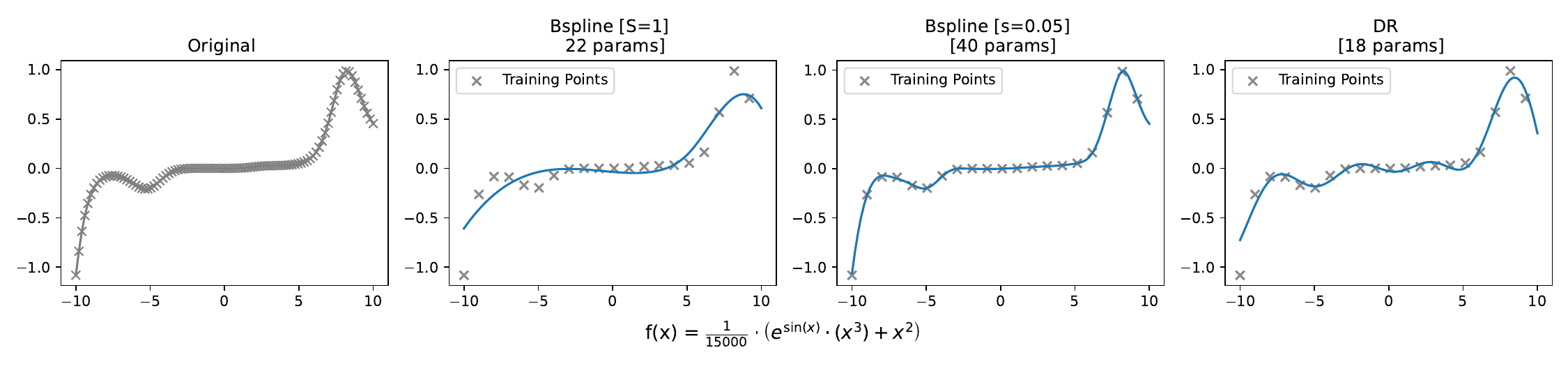}
    \caption{Parameter efficiency and function fitting capabilities of B-Splines vs Data Re-Uploading models. We compare against 2 B-Spline models with smoothness 1 and 0.05 resp. It is clear from the image that our DR circuit performs much better than the S=1 B-Spline and is slightly worse than the S=0.05 B-Spline.}
    \label{fig:parameter_efficiency}
\end{figure}

As we can see in Fig. \ref{fig:parameter_efficiency}, the \dr circuit is able to fit the function $\frac{1}{15000}\left(e^{\sin(x)}\cdot x^3 + x^2\right)$ using only 16 parameters. In comparison, even using 22 parameters B-Splines are not able to fit the function and only at 46 parameters do we see a good approximation. A larger set of comparisons can be found in Figure. \ref{fig:full_dr_vs_bspline}.

The increased efficiency of \quirk units translates to smaller more efficient models that are able to achieve similar performance to classical KANs. This is evident in Table \ref{tab:KAN_vs_qKAN} where we see that \quirk models are able to achieve similar RMSE values to classical KANs while using significantly fewer parameters.

\subsection{Interpretability of \quirk:} \label{subsection:Interpretability}
\begin{figure}[h]
    \centering
    \includegraphics[width=0.95\columnwidth]{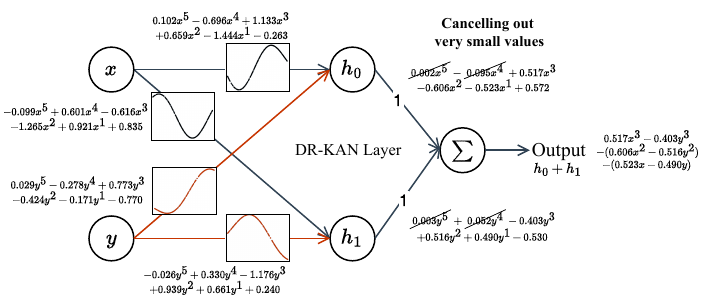}
    \caption{A single layer , 2 unit \quirk with a simple sum on the output approximating the function $f(x,y) = x^2 - y^2$. We can see that the outputs of the \dr units can be fit to 1-D polynomials. We can then analyze the polynomials to construct a simplified output.}
    \label{fig:Interpretability}
\end{figure}

\quirk models maintain the interpretability advantages of KANs. Due to still being composed of univariate functions, we can still analyze the outputs of each unit in the network. As seen in Fig. \ref{fig:Interpretability}, we can fit the outputs of each \dr unit to simple polynomials and then analyze these polynomials to construct a simplified output function. This allows us to derive closed form equations from trained \quirk models similar to KANs.

\section{Conclusion} \label{section:conclusion}
In this paper we have presented \quirk as an analogous and an efficient alternative to classical KANs. \quirk uses batches of single qubit DR models as univariate function approximators as opposed to B-splines. Due to the efficiency of single qubit DRs, the learnable activation functions could be represented in much fewer parameters as compared to B-splines. Our experiments show that DR-KAN can achieve loss values similar to KANs with much fewer parameter counts \ref{tab:KAN_vs_qKAN_full}. DR-KAN also outperforms standalone DR models which is conclusive from table \ref{tab:DR_KAN_vs_DR_full}.

\newpage
\bibliography{references}

\newpage
\appendix
\setcounter{figure}{0}
\setcounter{table}{0}
\setcounter{equation}{0}

\renewcommand{\thefigure}{A.\arabic{figure}}
\renewcommand{\thetable}{A.\arabic{table}}
\renewcommand{\theequation}{A.\arabic{equation}}

\newpage
\section{Appendix / Supplemental Material}
We include additional figures and tables to provide further insight into our experiments and results.

\subsection{Supplementary Figures}
\begin{figure}[h]
    \centering
    \includegraphics[width=0.98\textwidth]{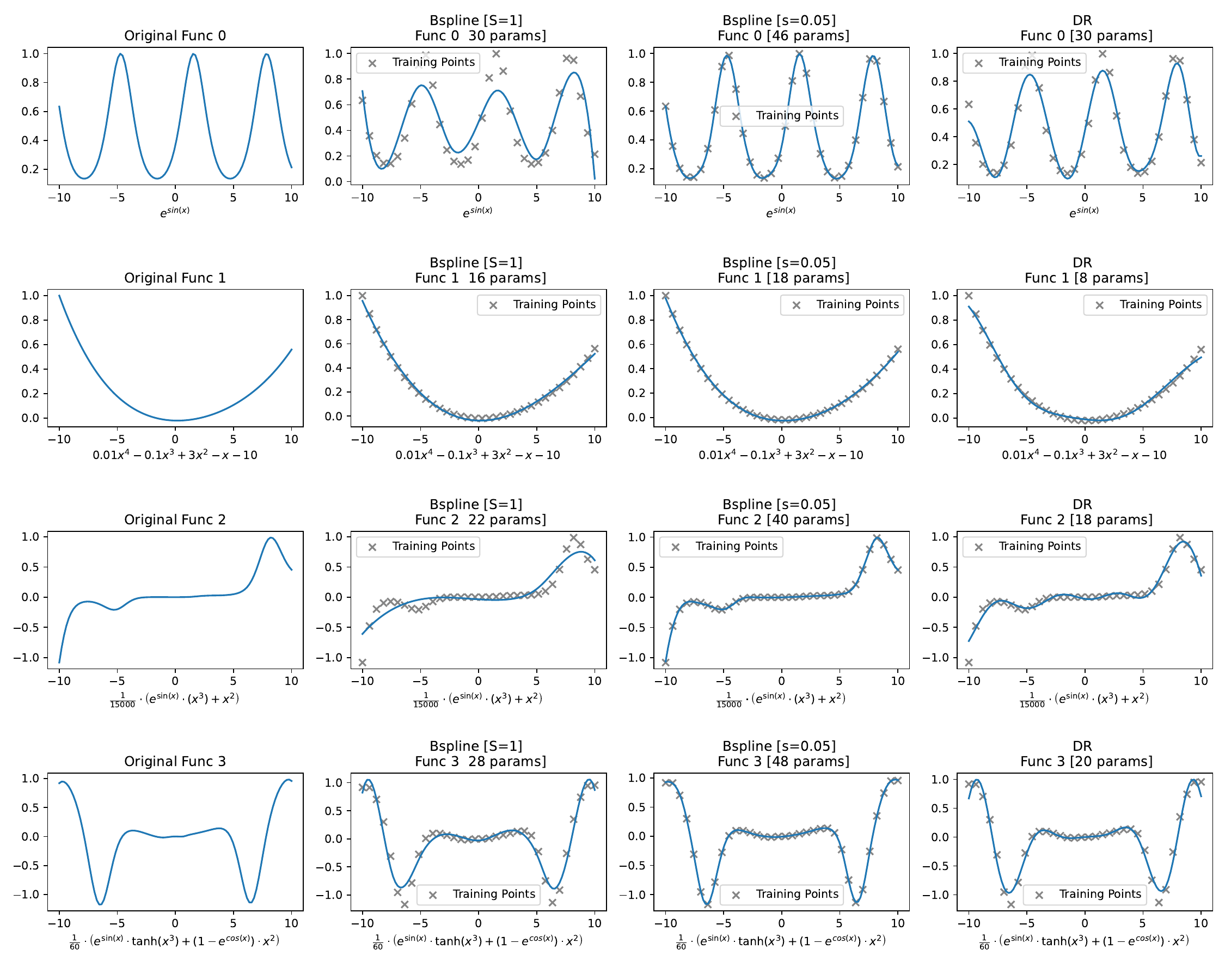}
    \caption{Expanded DR vs B-Spline Comparisons}
    \label{fig:full_dr_vs_bspline}
\end{figure}

\begin{figure}[t]
    \centering
    \includegraphics[width=0.65\textwidth, height=5.5cm]{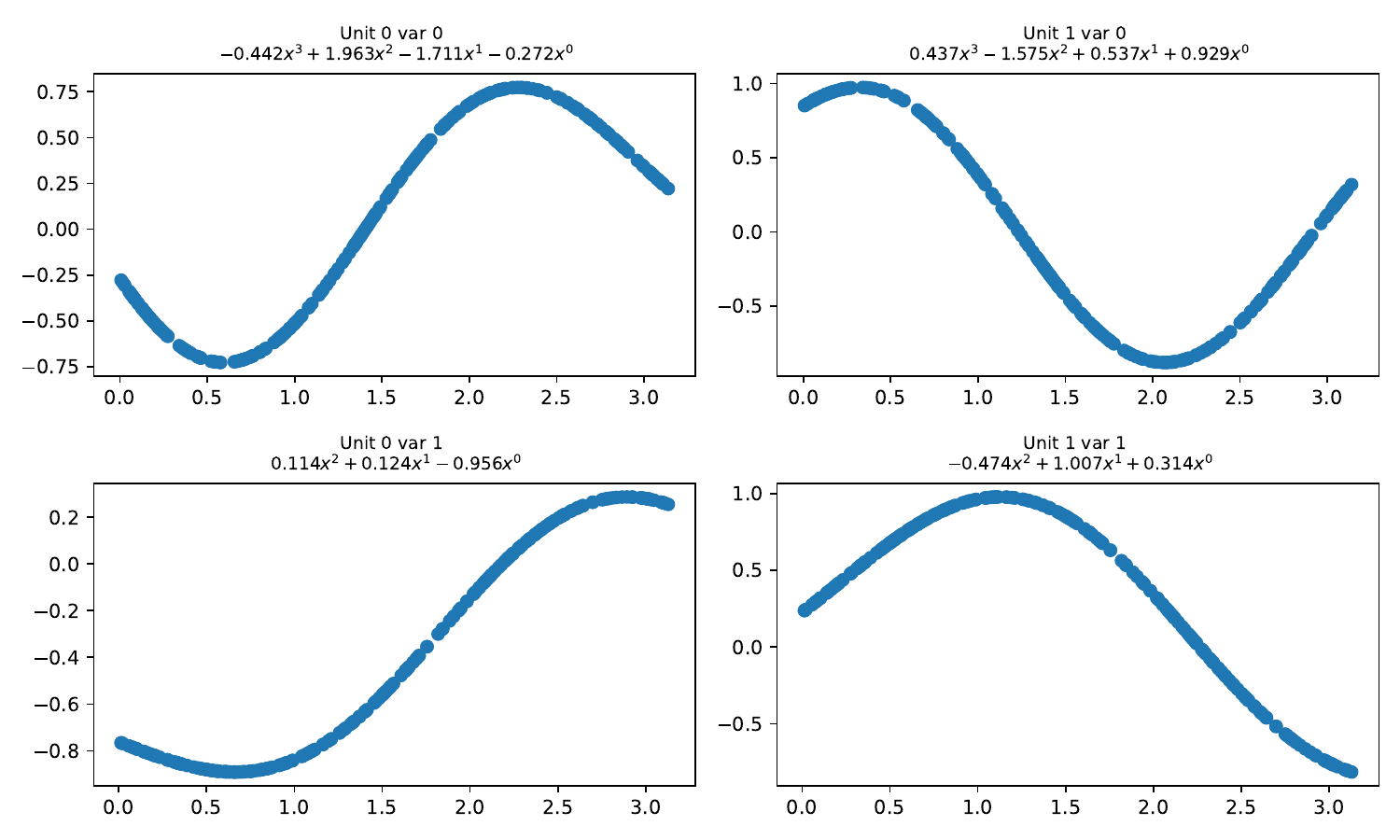}
    \caption{The functions learned by the \quirk model described in Section \ref{subsection:Interpretability} and the polynomials for the functions}
    \label{fig:full_Interpretability}
\end{figure}

\pagebreak

\subsection{Supplementary Tables}

\begin{table}[h]
    \centering
    \caption{Full Comparison between all \quirk and Classical KAN variants}
    \begin{tabular}{@{}l lc lc lc lc@{}}
        \toprule
        Equation  & \multicolumn{2}{c}{\quirk (Ours)} & \multicolumn{2}{c}{\quirk Pruned} & \multicolumn{2}{c}{Classical KAN} & \multicolumn{2}{c}{Classical KAN Pruned}
        \\ \cmidrule(lr){2-3} \cmidrule(lr){4-5} \cmidrule(lr){6-7} \cmidrule(lr){8-9}
                  & Loss                              & \# Params                         & Loss                              & \# Params                                & Loss                 & \# Params & Loss                 & \# Params \\ \midrule
        I.6.2     & $1.40\times 10^{-3}$              & 36                                & $2.95\times 10^{-2}$              & 24                                       & $3.90\times 10^{-1}$ & 136       & $1.51\times 10^{0}$  & 56        \\
        I.6.2b    & $2.64\times 10^{-3}$              & 72                                & $2.01\times 10^{-2}$              & 32                                       & $3.68\times 10^{-1}$ & 244       & $6.91\times 10^{-1}$ & 72        \\
        I.9.18    & $9.00\times 10^{-3}$              & 234                               & $1.06\times 10^{-2}$              & 120                                      & $4.99\times 10^{-1}$ & 736       & $1.65\times 10^{0}$  & 120       \\
        I.12.11   & $1.04\times 10^{-2}$              & 48                                & $5.64\times 10^{-2}$              & 24                                       & $9.90\times 10^{-4}$ & 136       & $1.58\times 10^{-1}$ & 56        \\
        I.13.12   & $4.20\times 10^{-3}$              & 36                                & $7.67\times 10^{-2}$              & 24                                       & $2.54\times 10^{1}$  & 136       & $2.50\times 10^{1}$  & 56        \\
        I.15.3x   & $1.30\times 10^{-2}$              & 36                                & $1.84\times 10^{-2}$              & 24                                       & $5.14\times 10^{-1}$ & 80        & $6.48\times 10^{-1}$ & 56        \\
        I.16.6    & $1.38\times 10^{-2}$              & 36                                & $6.50\times 10^{-2}$              & 24                                       & $3.34\times 10^{-2}$ & 136       & $4.40\times 10^{-1}$ & 56        \\
        I.18.4    & $8.00\times 10^{-3}$              & 36                                & $5.17\times 10^{-2}$              & 24                                       & $1.31\times 10^{1}$  & 136       & $4.87\times 10^{1}$  & 56        \\
        I.26.2    & $7.07\times 10^{-2}$              & 60                                & $2.81\times 10^{-2}$              & 40                                       & $4.20\times 10^{-3}$ & 136       & $1.71\times 10^{-1}$ & 56        \\
        I.27.6    & $2.31\times 10^{-2}$              & 54                                & $2.37\times 10^{-2}$              & 36                                       & $3.25\times 10^{-1}$ & 136       & $1.28\times 10^{0}$  & 56        \\
        I.29.16   & $1.24\times 10^{-2}$              & 51                                & $8.99\times 10^{-2}$              & 35                                       & $7.81\times 10^{-3}$ & 391       & $1.99\times 10^{-1}$ & 72        \\
        I.30.3    & $4.79\times 10^{-2}$              & 43                                & $7.05\times 10^{-2}$              & 27                                       & $4.13\times 10^{-3}$ & 136       & $4.43\times 10^{-3}$ & 56        \\
        I.30.5    & $5.53\times 10^{-2}$              & 82                                & $6.37\times 10^{-2}$              & 40                                       & $3.31\times 10^{-1}$ & 156       & $4.92\times 10^{-1}$ & 56        \\
        I.37.4    & $7.50\times 10^{-2}$              & 76                                & $1.56\times 10^{-1}$              & 40                                       & $9.08\times 10^{-3}$ & 136       & $4.82\times 10^{-2}$ & 56        \\
        I.40.1    & $8.73\times 10^{-2}$              & 68                                & $5.85\times 10^{-2}$              & 40                                       & $3.65\times 10^{-2}$ & 136       & $3.88\times 10^{-1}$ & 56        \\
        I.44.4    & $3.74\times 10^{-2}$              & 48                                & $3.61\times 10^{-2}$              & 24                                       & $4.63\times 10^{-2}$ & 156       & $6.26\times 10^{-1}$ & 56        \\
        I.50.26   & $2.40\times 10^{-2}$              & 54                                & $4.86\times 10^{-2}$              & 36                                       & $4.10\times 10^{-4}$ & 192       & $5.19\times 10^{-2}$ & 56        \\
        II.2.42   & $9.98\times 10^{-2}$              & 36                                & $6.49\times 10^{-2}$              & 24                                       & $3.90\times 10^{-4}$ & 136       & $3.33\times 10^{-1}$ & 56        \\
        II.6.15a  & $2.04\times 10^{-2}$              & 76                                & $4.60\times 10^{-2}$              & 52                                       & $2.00\times 10^{-3}$ & 202       & $1.31\times 10^{-2}$ & 72        \\
        II.11.7   & $8.55\times 10^{-2}$              & 72                                & $6.68\times 10^{-2}$              & 48                                       & $8.21\times 10^{-3}$ & 272       & $1.98\times 10^{-1}$ & 84        \\
        II.11.27  & $4.92\times 10^{-2}$              & 36                                & $5.93\times 10^{-2}$              & 24                                       & $4.48\times 10^{-2}$ & 136       & $2.20\times 10^{-1}$ & 56        \\
        II.35.18  & $5.24\times 10^{-2}$              & 20                                & $6.59\times 10^{-2}$              & 14                                       & $1.26\times 10^{1}$  & 136       & $1.17\times 10^{1}$  & 56        \\
        II.36.38  & $2.72\times 10^{-3}$              & 51                                & $1.32\times 10^{-2}$              & 35                                       & $5.40\times 10^{-4}$ & 244       & $3.35\times 10^{-1}$ & 72        \\
        II.38.3   & $6.10\times 10^{-3}$              & 72                                & $4.79\times 10^{-2}$              & 32                                       & $2.54\times 10^{1}$  & 80        & $2.33\times 10^{1}$  & 56        \\
        III.9.52  & $3.06\times 10^{-2}$              & 102                               & $4.97\times 10^{-2}$              & 62                                       & $2.20\times 10^{-4}$ & 342       & $8.93\times 10^{-2}$ & 72        \\
        III.10.19 & $1.41\times 10^{-2}$              & 58                                & $6.55\times 10^{-2}$              & 40                                       & $3.00\times 10^{-5}$ & 80        & $3.64\times 10^{-3}$ & 56        \\
        III.17.37 & $1.11\times 10^{-3}$              & 76                                & $7.52\times 10^{-2}$              & 52                                       & $3.43\times 10^{-3}$ & 244       & $2.81\times 10^{-1}$ & 72        \\
        \bottomrule
    \end{tabular}
    \label{tab:KAN_vs_qKAN_full}
\end{table}

\begin{table}[h]
    \centering
    \caption{Full Comparison between all DR-KAN \quirk  variants and Vanilla DR Models}
    \begin{tabular}{@{}l lc lc lc@{}}
        \toprule
        Equation  & \multicolumn{2}{c}{\quirk (Ours) Pruned} & \multicolumn{2}{c}{Vanilla DR}
        \\ \cmidrule(lr){2-3} \cmidrule(lr){4-5}
                  & Loss                                     & \# Params                      & Loss                 & \# Params \\ \midrule
        I.6.2     & $2.95\times 10^{-2}$                     & 24                             & $1.46\times 10^{-1}$ & 30        \\
        I.6.2b    & $2.01\times 10^{-2}$                     & 32                             & $6.65\times 10^{-1}$ & 30        \\
        I.9.18    & $1.06\times 10^{-2}$                     & 120                            & $2.07\times 10^{1}$  & 30        \\
        I.12.11   & $5.64\times 10^{-2}$                     & 24                             & $1.48\times 10^{-1}$ & 30        \\
        I.13.12   & $7.67\times 10^{-2}$                     & 24                             & $3.13\times 10^{1}$  & 30        \\
        I.15.3x   & $1.84\times 10^{-2}$                     & 24                             & $3.64\times 10^{0}$  & 30        \\
        I.16.6    & $6.50\times 10^{-2}$                     & 24                             & $2.89\times 10^{-1}$ & 30        \\
        I.18.4    & $5.17\times 10^{-2}$                     & 24                             & $2.83\times 10^{-1}$ & 30        \\
        I.26.2    & $2.81\times 10^{-2}$                     & 40                             & $1.67\times 10^{-1}$ & 30        \\
        I.27.6    & $2.37\times 10^{-2}$                     & 36                             & $6.52\times 10^{-2}$ & 30        \\
        I.29.16   & $8.99\times 10^{-2}$                     & 35                             & $8.03\times 10^{-1}$ & 30        \\
        I.30.3    & $7.05\times 10^{-2}$                     & 27                             & $1.01\times 10^{0}$  & 30        \\
        I.30.5    & $6.37\times 10^{-2}$                     & 40                             & $2.88\times 10^{-1}$ & 30        \\
        I.37.4    & $1.56\times 10^{-1}$                     & 40                             & $3.36\times 10^{-1}$ & 30        \\
        I.40.1    & $5.85\times 10^{-2}$                     & 40                             & $1.38\times 10^{-1}$ & 30        \\
        I.44.4    & $3.61\times 10^{-2}$                     & 24                             & $2.49\times 10^{0}$  & 30        \\
        I.50.26   & $4.86\times 10^{-2}$                     & 36                             & $3.57\times 10^{-1}$ & 30        \\
        II.2.42   & $6.49\times 10^{-2}$                     & 24                             & $1.68\times 10^{0}$  & 30        \\
        II.6.15a  & $4.60\times 10^{-2}$                     & 52                             & $3.89\times 10^{0}$  & 30        \\
        II.11.7   & $6.68\times 10^{-2}$                     & 48                             & $3.66\times 10^{0}$  & 30        \\
        II.11.27  & $5.93\times 10^{-2}$                     & 24                             & $2.56\times 10^{2}$  & 30        \\
        II.35.18  & $6.59\times 10^{-2}$                     & 14                             & $2.09\times 10^{-1}$ & 30        \\
        II.36.38  & $6.32\times 10^{-2}$                     & 35                             & $2.36\times 10^{0}$  & 30        \\
        II.38.3   & $4.79\times 10^{-2}$                     & 32                             & $1.35\times 10^{2}$  & 30        \\
        III.9.52  & $4.97\times 10^{-2}$                     & 62                             & $9.12\times 10^{-2}$ & 30        \\
        III.10.19 & $6.55\times 10^{-2}$                     & 40                             & $5.34\times 10^{-1}$ & 30        \\
        III.17.37 & $7.52\times 10^{-2}$                     & 52                             & $3.35\times 10^{0}$  & 30        \\
    \bottomrule
    \end{tabular}
    \label{tab:DR_KAN_vs_DR_full}
\end{table}

\begin{table*}[ht]
\centering
\caption{Performance comparison of equations with DR-KAN, KAN and MLP models. Losses in RMSE. Results of KAN and MLP (Best Accuracy) directly from KAN paper. Our Shape is [2,1] with DR layers [3,3] for 43 params and [4,3] for 51 params.}
\begin{tabular}{lcccccccccccc}
\toprule
\textbf{Equation} & \textbf{DR-KAN} & \textbf{\# Params} & \textbf{Min KAN}& \textbf{Min KAN} & \textbf{Best KAN} & \textbf{Best KAN} & \textbf{MLP} \\
\midrule
JacobElliptic       & $8.78 \times 10^{-3}$ & 55 & [2,2,1]       & $7.29 \times 10^{-3}$ & [2,3,2,1,1,1]   & $\mathbf{1.33 \times 10^{-4}}$ & $6.48 \times 10^{-4}$ \\
EllipticInt1      & $2.79 \times 10^{-3}$ & 43 & [2,2,1]       & $1.00 \times 10^{-3}$ & [2,2,1,1,1]     & $\mathbf{1.24 \times 10^{-4}}$ & $5.52 \times 10^{-4}$ \\
EllipticInt2      & $1.26 \times 10^{-2}$ & 43 & [2,2,1,1]     & $8.36 \times 10^{-5}$ & [2,2,1,1]       & $\mathbf{8.26 \times 10^{-5}}$ & $3.04 \times 10^{-4}$ \\
BesselJ                & $3.95 \times 10^{-2}$ & 43 & [2,2,1]       & $4.93 \times 10^{-3}$ & [2,3,1,1,1]     & $\mathbf{1.64 \times 10^{-3}}$ & $5.52 \times 10^{-3}$ \\
BesselY                & $2.59 \times 10^{-2}$ & 43  & [2,3,1]       & $1.89 \times 10^{-3}$ & [2,2,2,1]       & $\mathbf{1.49 \times 10^{-3}}$ & $3.45 \times 10^{-4}$ \\
BesselK                & $2.48 \times 10^{-2}$ & 43  & [2,1,1]       & $4.89 \times 10^{-3}$ & [2,2,1]         & $\mathbf{2.52 \times 10^{-5}}$ & $1.67 \times 10^{-4}$ \\
BesselI                & $\mathbf{4.12 \times 10^{-3}}$                  & 43  & [2,4,3,2,1,1] & $9.28 \times 10^{-3}$ & [2,4,3,2,1,1]   & $9.28 \times 10^{-3}$ & $1.07 \times 10^{-2}$ \\
Legendre0              & $2.66 \times 10^{-3}$ & 43  & [2,2,1]       & $5.25 \times 10^{-5}$ & [2,2,1]         & $\mathbf{5.25 \times 10^{-5}}$ & $1.74 \times 10^{-2}$ \\
Legendre1              & $7.22 \times 10^{-3}$                   & 51 & [2,4,1]       & $6.90 \times 10^{-4}$ & [2,4,1]         & $\mathbf{6.90 \times 10^{-4}}$ & $1.50 \times 10^{-3}$ \\
Legendre2              & $8.66 \times 10^{-3}$                   & 43 & [2,2,1]       & $4.88 \times 10^{-3}$ & [2,3,2,1]       & $\mathbf{2.26 \times 10^{-3}}$ & $9.43 \times 10^{-4}$ \\
SH01    & $5.16 \times 10^{-3}$ & 43 & [2,1,1]       & $2.21 \times 10^{-7}$ & [2,1,1]         & $\mathbf{2.21 \times 10^{-7}}$ & $1.25 \times 10^{-6}$ \\
SH11    & $2.84 \times 10^{-2}$ & 43 & [2,2,1]       & $7.86 \times 10^{-4}$ & [2,3,2,1]       & $\mathbf{1.22 \times 10^{-4}}$ & $6.70 \times 10^{-4}$ \\
SH02    & $8.29 \times 10^{-3}$ & 43 & [2,1,1]       & $1.95 \times 10^{-7}$ & [2,1,1]         & $\mathbf{1.95 \times 10^{-7}}$ & $2.85 \times 10^{-6}$ \\
SH12    & $2.09 \times 10^{-2}$ & 43 & [2,2,1]       & $4.70 \times 10^{-4}$ & [2,2,1,1]       & $\mathbf{1.50 \times 10^{-5}}$ & $1.84 \times 10^{-3}$ \\
SH22    & $6.44 \times 10^{-2}$ & 43 & [2,2,1]       & $1.12 \times 10^{-3}$ & [2,2,3,2,1]     & $\mathbf{9.45 \times 10^{-5}}$ & $6.21 \times 10^{-4}$ \\
\bottomrule
\end{tabular}
\end{table*}

\begin{table}[h!]
\centering
\caption{RMSE Loss for MLPs and KAN on the Feynman dataset when the outputs are constrained to [-1,1].}
\begin{tabular}{lcccccc}
\toprule
& \multicolumn{4}{c}{MLPs}& \multicolumn{2}{c}{DR-KAN}\\
\midrule
& \multicolumn{2}{c}{Less Params}& \multicolumn{2}{c}{Sufficient Params}\\
\midrule
Equation & Loss & \# Params & Loss & \# Params & Loss (Ours) & \# Params \\
\midrule
1.6.2   & $5.69 \times 10^{-3}$ & 133 & $1.43 \times 10^{-3}$ & 2305 & $1.40 \times 10^{-3}$ & 36 \\
1.6.2b  & $4.61 \times 10^{-3}$ & 57  & $6.16 \times 10^{-3}$ & 253  & $2.64 \times 10^{-3}$ & 72 \\
1.9.18  & $3.42 \times 10^{-3}$ & 133 & $3.28 \times 10^{-4}$ & 2305 & $9.00 \times 10^{-3}$ & 234 \\
1.12.11 & $3.33 \times 10^{-3}$ & 57  & $3.96 \times 10^{-4}$ & 565  & $1.40 \times 10^{-2}$ & 48 \\
1.13.12 & $2.21 \times 10^{-3}$ & 133 & $2.36 \times 10^{-4}$ & 565  & $4.20 \times 10^{-3}$ & 36 \\
1.15.3x & $9.35 \times 10^{-3}$ & 133 & $4.79 \times 10^{-4}$ & 565  & $1.30 \times 10^{-2}$ & 36 \\
1.16.6  & $1.61 \times 10^{-2}$ & 133 & $2.60 \times 10^{-3}$ & 565  & $1.38 \times 10^{-2}$ & 36 \\
1.18.4  & $6.07 \times 10^{-3}$ & 133 & $7.59 \times 10^{-4}$ & 565  & $8.00 \times 10^{-3}$ & 36 \\
1.26.2  & $4.02 \times 10^{-3}$ & 133 & $1.66 \times 10^{-3}$ & 2305 & $7.07 \times 10^{-3}$ & 60 \\
1.29.16 & $2.47 \times 10^{-3}$ & 133 & $1.02 \times 10^{-3}$ & 2305 & $2.31 \times 10^{-2}$ & 54 \\
1.30.2  & $4.00 \times 10^{-3}$ & 133 & $1.01 \times 10^{-3}$ & 565  & $1.24 \times 10^{-2}$ & 51 \\
1.37.6  & $8.97 \times 10^{-2}$ & 133 & $2.19 \times 10^{-3}$ & 2305 & $4.79 \times 10^{-3}$ & 50 \\
1.44.4  & $3.10 \times 10^{-3}$ & 133 & $4.18 \times 10^{-4}$ & 2305 & $5.53 \times 10^{-3}$ & 82 \\
1.44.6  & $3.29 \times 10^{-3}$ & 133 & $3.12 \times 10^{-4}$ & 565  & $7.50 \times 10^{-3}$ & 76 \\
II.4.2  & $5.62 \times 10^{-4}$ & 133 & $1.02 \times 10^{-4}$ & 565  & $8.73 \times 10^{-3}$ & 60 \\
II.6.15a& $1.01 \times 10^{-3}$ & 133 & $9.30 \times 10^{-5}$ & 2305 & $5.00 \times 10^{-3}$ & 48 \\
II.11.7 & $9.47 \times 10^{-4}$ & 133 & $4.55 \times 10^{-4}$ & 2305 & $2.40 \times 10^{-3}$ & 54 \\
II.15.18& $1.01 \times 10^{-3}$ & 133 & $3.93 \times 10^{-4}$ & 565  & $9.98 \times 10^{-3}$ & 36 \\
II.35.36a&$1.24 \times 10^{-3}$ & 133 & $8.30 \times 10^{-5}$ & 565  & $2.04 \times 10^{-3}$ & 48 \\
II.38.7 & $2.15 \times 10^{-3}$ & 133 & $2.33 \times 10^{-4}$ & 565  & $8.55 \times 10^{-3}$ & 72 \\
II.9.52 & $2.65 \times 10^{-3}$ & 133 & $1.43 \times 10^{-4}$ & 565  & $4.92 \times 10^{-3}$ & 36 \\
II.10.19& $1.99 \times 10^{-3}$ & 133 & $3.00 \times 10^{-4}$ & 2305 & $5.24 \times 10^{-3}$ & 20 \\
III.17.37&$1.63 \times 10^{-3}$ & 133 & $1.52 \times 10^{-4}$ & 2305 & $2.72 \times 10^{-3}$ & 36 \\
\bottomrule
\end{tabular}
\end{table}

\end{document}